\documentclass{pasj00}
\draft

\begin{document}
\SetRunningHead{Author(s) in page-head}{Running Head}
\Received{????/??/??}
\Accepted{????/??/??}

\title{The Differences of Star Formation History Between
Merging Galaxies and Field Galaxies in the EDR of the SDSS}

\author{Zhi-Jian \textsc{Luo}
}
\affil{Department of Physics, Jiangxi Normal
University, Nanchang 330022, China} \email{zjluo@center.shao.ac.cn}

\author{Cheng-Gang \textsc{Shu}}
\affil{Joint Center for Astrophysics, Shanghai Normal University,
Shanghai 200234, China \\Shanghai Astronomical Observatory, Chinese
Academy of Sciences, Shanghai 200030,
China}\email{cgshu@center.shao.ac.cn} \and
\author{Jia-Sheng {\sc Huang }}
\affil{Harvard-Smithsonian Center for Astrophysics, 60 Garden
Street, Cambridge, MA 02138, USA}\email{jhuang@cfa.harvard.edu}

%


\maketitle

\begin{abstract}
Based on the catalog of merging galaxies in the Early Data Release
(EDR) of the Sloan Digital Sky Survey  (SDSS), the differences of
star formation history between merging galaxies and field galaxies
are studied statistically by means of three spectroscopic indicators
the 4000-\r{A} break strength, the Balmer absorption-line index, and
the specific star formation rate. It is found that for early-type
merging galaxies the interactions will not induce significant
enhancement of the star-formation activity because of its stability
and lack of cool gas. On the other hand, late-type merging galaxies
always in general display more active star formation than field
galaxies on different timescales within about 1Gyr. We also conclude
that the mean stellar ages of late-type merging galaxies are younger
than those of late-type field galaxies.
\end{abstract}

\section{Introduction}

According to the current hierarchical
galaxy formation theory, galaxy merging and interacting play crucial
roles in determining their properties and interpreting certain
unusual phenomena in galaxies (e.g. the starbursts and the
occurrence of activity galaxy nucleus). Observations also show that
merging and interacting galaxies often have special morphologies,
such as galactic bridges and tidal tails. Moreover, they generally
display stronger emission in H$\alpha$, radio continuum and infrared
than field galaxies(\cite{kr87}; \cite{bu87}; \cite{hu81};
\cite{co82}; \cite{lo84}; \cite{so88}; \cite{jo84}), which implies
that galaxy interaction can always trigger strong star formation
activities.

Numerical simulations have been widely adopted to study how galaxy
interaction and merging proceed star formation activity, and many
useful results have been obtained (\cite{tt72}; \cite{mh96}). On one
hand it is hard to give the detailed descriptions theoretically
about star formation history due to finite resolution in simulations
and complicated physical processes of inter-stellar medium (ISM).
From observational point of view, on the other hand, different
samples of interacting galaxies have been studied statistically
galaxy interaction on star formation since many years ago
(\cite{kr87}; \cite{lt78}; \cite{bw88}; \cite{bl03}), while their
results cannot be in good consistence with each other: some studies
showed that galaxy interactions have enhanced star formation
activity greatly and some did not. Note that inconsistent results
are also shown recently by Lambas et al. (2003) and Nikolic et al.
(2004), respectively, according to their more reliable statistic
results based on thousands of galaxy pairs. In fact, the connection
between star formation and galaxy interaction has been suspected
ever since the first large survey was available.

In this paper we investigate statistically this problem according to
the catalog of merging galaxies in the Early Data Release (EDR) of
the Sloan Digital Sky Survey (SDSS) (\cite{yo00}; \cite{st02};
\cite{ho01}; \cite{sr02}) provided by Allam et al. (2004). This
catalog is deeper in redshift and wider in the sky area than most of
the other wide-area catalog up to date. Same as Kauffmann et al.
(2004), three spectroscopic indicators, the 4000-\r{A} break
strength D$_{n}(4000)$, the Balmer absorption-line index
H$\delta_{A}$ and the specific star formation rate (SFR) SFR/$M_{*}$
with $M_{*}$ the stellar masses in a galaxy, are adopted as
diagnostics to study the star formation histories of galaxies. The
reasons are as follows. Since SFR/$M_{*}$ of individual galaxies are
estimated by the nebular emission lines (particularly the H$\alpha$
line), it probes star formation activity on a timescale similar to
the lifetime of a typical HII region, i.e., $\sim 10^{7}$ years. The
index H$\delta_{A}$ of the absorption line denotes the timescale of
$3\times10^{8}$ years because it peaks once hot O and B stars have
terminated their evolution and the optical light is dominated by
late-B to early F-stars, while the strength of the 4000-\r{A} break
D$_{n}(4000)$ increases monotonically with time (but at stellar ages
of more than $\sim$1Gyr, metallicity effects also become important)
(\cite{kw04}; Kauffmann et al. 2003a,b; Goto 2005, 2004;
\cite{go03}).


\section{the Sample and data}

The SDSS is a digital photometric and spectroscopic survey that
covers one-quarter of the entire sky by using a dedicated 2.5-meter
wide-field telescope at the Apache Point Observatory(\cite{yo00};
\cite{st02}). The EDR of SDSS consists of 462 square degrees of
imaging data in $u$, $g$, $r$, $i$ and $z$  bands together with
medium-resolution spectra for approximately 40,000 galaxies and 4000
quasars. By the SDSS imaging reduction software (PHOTO)
(\cite{li02}), Allam et al. (2004) identified isolated merging
galaxy pair candidates for galaxies in the magnitude range 16.0
$\leq$ $g^{*}$ $\leq$ 21.0 within the EDR through their automated
systematic search routine. In their routine, a merging pair is
basically defined as one in which the centers of two galaxies has
the separation less than the sum of the members' Petrosian radii.
Their selection algorithm implemented a variation on the original
Karachentsev(1972)'s isolated pair criteria and proved to be very
efficient and fast. Moreover, to remove spurious pairs due to poor
image deblending, Allam et al. (2004) also inspected all the merging
pairs by eyes. After all rejections and verifications the final
number of candidates of merging pairs is 1479, most of which have
only one member galaxy with known spectroscopic redshift. And Allam
et al (2004) also estimated that the contamination by chance
projections is less than 3.4\%.

There are 744 galaxies with known spectroscopic redshifts among
these 1479 merging galaxy pairs. Based on all available
spectroscopic observations in the EDR of the SDSS, it is found that
581 of the above 744 merging galaxies have the spectroscopic
indicators D$_{n}(4000)$, H$\delta_{A}$ and SFR/$M_{*}$ measured by
Kauffmann et al. (2004), who obtained these indicators according to
a special-purpose code described in detail by Tremonti et al (2004).
We take these 581 merging galaxies as our sample for further studies
below.

Since numerical simulations of pre-prepared mergers showed that
interactions between axis-symmetrical systems without bulges or with
small ones might induce gas inflowing to the central region of the
systems and triggering starburst episodes while the interactions
would not trigger strong starburst for galaxies with dense bulges
before they finally collide due to their deep potential wells, we
divide these 581 selected galaxies into two sub-samples: the
early-type galaxy sub-sample (hereafter EGS) and the late-type
galaxy sub-sample (hereafter LGS) by adopting a galaxy morphological
classification scheme of the concentration parameter $c$ (defined as
the radio of Petrosian $90\%$- to $50\%$-light radii as measure in
the r-band, i.e. $c=R_{90}/R_{50}$) among member galaxies. According
to Shimasaku et al. (2001) and Strateva et al. (2001), $c \sim 2.6$
is adopted in the present paper as the selection criterion for
early- and late-type galaxies and the resulted numbers of galaxies
in the EGS and LGS are 336 and 245 respectively.

 There are 21 merging galaxy pairs within these selected 581
galaxies. The numbers of pairs with two early-type galaxies and two
late-type galaxies are 13 and 5 respectively. Only 3 pairs show
their members as one early-type galaxy and another late-type galaxy.
So, the contamination of the above classification for the EGS or the
LGS in studying the star formation history of merging pairs below is
very small.

To investigate the effect of interaction on star formation histories
in merging galaxies, we construct the corresponding samples of field
galaxies for individual sub-samples of merging galaxies respectively
by a Monte Carlo algorithm from the full EDR spectroscopic galaxy
catalog, which is named as the control galaxy samples (hereafter
CGSs). Each CGS has the same number of galaxies as the corresponding
sub-samples of merging galaxies. Comparing the EGS and the LGS with
their corresponding CGSs respectively, we can investigate the
differences between them and unveil the effects of the interactions.

Since the star formation histories of galaxies are strongly
correlated with their stellar mass, concentration parameter and
redshift(Kauffman et al. 2003a,b), the influences of these factors
must be eliminated when we compare the EGS and LGS with their
corresponding CGSs, which is as follows. Galaxies in CGSs of
individual corresponding merging galaxies with known spectroscopic
redshifts and measured spectroscopic indicators D$_{n}(4000)$,
H$\delta_{A}$ and SFR/$M_{*}$ are selected to have the similar
redshifts, stellar masses and concentration parameter from the full
EDR spectroscopic galaxy catalog. The distributions of redshifts,
stellar masses and concentration parameters in CGSs mirror those of
the EGS and LGS. It should be pointed out that the random
constructing samples (CGSs) are approximates the samples of field
galaxies, since only about 10$\%$ of all galaxies lie in rich
clusters(\cite{at04}).

It is important to consider the effect of AGNs in star formation
history analysis. Same as Kauffmann et al. (2003), AGNs were
identified by their positions on the [NII]/H$\alpha$ vs
[OIII]/H$\beta$ planes. It is found that less than 18\% and less
than 8\% of galaxies in the EGS and LGS are identified as galaxies
with AGN features respectively. This implies that the contaminations
of AGNs in the present study do not significantly influence our
statistical results.

\section{The Differences of Star Formation History}

By comparing the distributions of three spectroscopic indicators
(D$_{n}(4000)$, H$\delta_{A}$ and SFR/$M_{*}$) of galaxies between
EGS and its corresponding CGS, no significant differences have been
found.  It is because that the interaction will not induce
significant enhancement of the star-formation activities for
galaxies in EGS before they finally collide according to the
previous numerical simulation (\cite{mh96}), since dense bulges of
early-type galaxies act to stabilize galaxies against the gas
inflows and the early-type galaxies lack of cool gas intrinsically.

Galaxies in the LGS and its corresponding CGS display different
properties of D$_{n}(4000)$, H$\delta_{A}$ and SFR/$M_{*}$ from
those in EGS and its corresponding CGS. As an example, the
distributions of the 4000-\r{A} break strength D$_{n}(4000)$ for
galaxies in the LGS and its corresponding CGS are plotted as the
solid and dashed histograms respectively in Fig. 1
Error bars in the figure for merging galaxies are estimated by
applying the bootstrap resampling technique (1000 random samples)
and those for CGS galaxies are taken as their standard Gaussian
deviations. One can find significant differences from the figure
that the median values of D$_{n}(4000)$ are about 1.3 and 1.5
respectively for galaxies in the LGS and the corresponding CGS.
Since strength of the 4000-\r{A} break increases monotonically with
time and it is an excellent age indicator for young stellar
population ($<1 \rm Gyr$) as mentioned above, the smaller
D$_{n}(4000)$ for merging galaxies implies that their average
stellar ages are younger than the corresponding field galaxies,
i.e., the merging galaxies in the LGS show stronger star formation
activities recently.

\begin{figure}
  \begin{center}
    \FigureFile(80mm,80mm){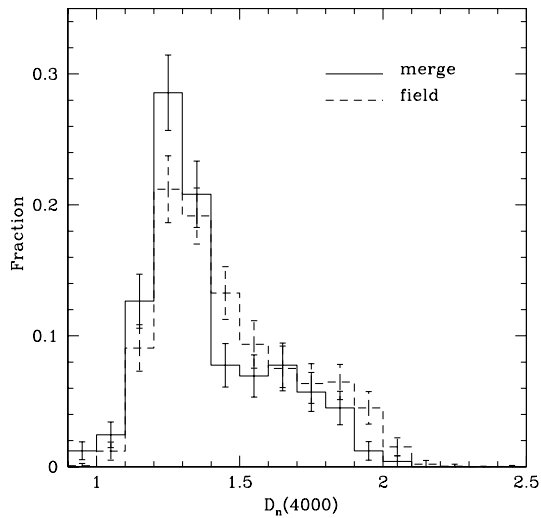}
  \end{center}
  \caption{The distributions of the 4000-\r{A} break strength
D$_{n}(4000)$ of merging galaxies(solid line) in the LGS and its
corresponding field galaxies(dashed line) in the CGS,
respectively.}\label{fig:fig1}
\end{figure}

The distributions of the Balmer absorption line index H$\delta_{A}$
for these two samples are plotted respectively in Fig. 2 with the
same notations as Fig. 1. From this figure one can also see that the
median values of H$\delta_{A}$ are about 4\r{A} and 3\r{A} for
galaxies in the LGS and the corresponding CGS respectively, i.e.,
merging galaxies in the LGS tend to have larger H$\delta_{A}$.
Because the strength of H$\delta_{A}$ absorption line peaks at about
$3\times10^{8}$ years after an episode of star formation, the larger
H$\delta_{A}$ indicates  that the merging galaxies display more
possibly a burst of star formation $0.1-1$Gyr ago.

\begin{figure}
  \begin{center}
    \FigureFile(80mm,80mm){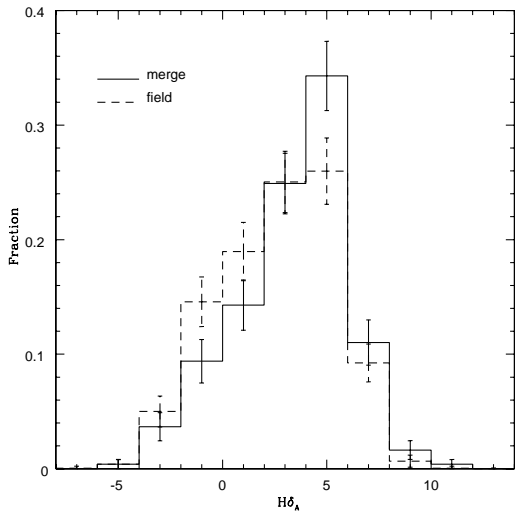}
  \end{center}
  \caption{The
distributions of the Balmer absorption line index H$\delta_{A}$ in
the LGS and its corresponding CGS, respectively, with the same
notations as Fig. 1.}\label{fig:fig2}
\end{figure}

Kauffmann et al. (2003) developed a method to distinguish recent
star formation histories dominated by bursts from those that are
more continuous. They pointed out that together with D$_{n}(4000)$
and H$\delta_{A}$ indices of a galaxy could allow one to constrain
the fraction of its stellar mass formed in the recent burst and the
mean age of the stellar population. Their results can be simply
summarized as follows. There are four regions in the
D$_{n}(4000)$-H$\delta_{A}$ plane. Two of them are the so-call
``starburst region". Galaxies located in ``starburst region" with
the onset of the burst occurred less than 0.1 Gyr ago always show
D$_{n}(4000)\lesssim1.1$; Galaxies with the onset of the burst
occurred more than 0.1 Gyr ago show the stronger H$\delta_{A}$ and
have $1.5\gtrsim$D$_{n}(4000)\gtrsim1.1$. This region is called as
the ``post-starburst region". The distributions of galaxies in the
D$_{n}(4000)$-H$\delta_{A}$ plane for our LGS and its corresponding
CGS are shown in Fig. 3. From the figure we can find that about
4$\%$ galaxies in the LGS locate in the 'starburst region', i.e.,
D$_{n}(4000)\lesssim1.1$ while less than 1$\%$ galaxies in the
corresponding CGS locate in this region. Moreover, it can be also
found that about 10\% galaxies in the LGS locate in the region with
$1.5\gtrsim$D$_{n}(4000)\gtrsim1.1$ and H$\delta_{A}\gtrsim6.0$\r{A}
while in the CGS the fraction is about 7\% which implies that more
galaxies in the LGS display 'post-starburst' phase.

\begin{figure}
  \begin{center}
    \FigureFile(80mm,80mm){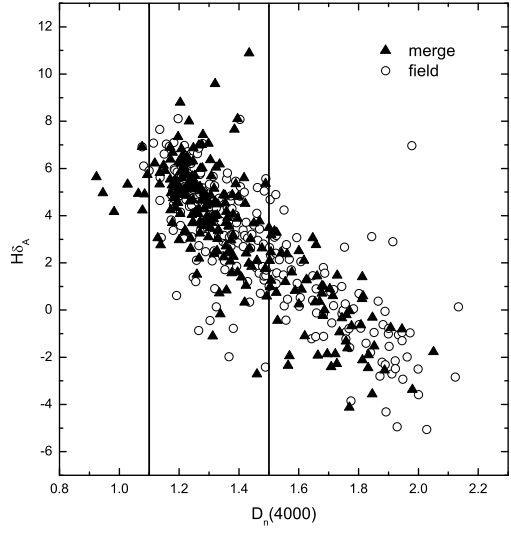}
  \end{center}
  \caption{The distribution of galaxies in the D$_{n}(4000)$-H$\delta_{A}$ plane.
  Solid triangles and open circles indicate the merging galaxies in the LGS
  and the
  field galaxies in the CGS. Vertical lines indicate D$_{n}(4000)=1.1$ (left) and
  D$_{n}(4000)=1.5$ (right).
    }\label{fig:fig3}
\end{figure}

Furthermore, the relative distributions of spectroscopic indicator
SFR/$M_{*}$ for galaxies of the LGS and its corresponding CGS are
shown respectively in Fig. 4 with the same notation as those in
Figs. 1 and 2. It is found that the distribution of galaxies in the
LGS (solid line) shifts to the higher SFR/$M_{*}$ than that in the
CGS (dashed line), which implies that the specific star formation
rates of merging galaxies are generally larger than those of field
galaxies. Since SFR/$M_{*}$ of a galaxy probes its star formation on
a timescale about $\sim 10^{7}$ years, this result also indicates
that merging galaxies have higher star formation activities at
present or not a long time ago.

\begin{figure}
  \begin{center}
    \FigureFile(80mm,80mm){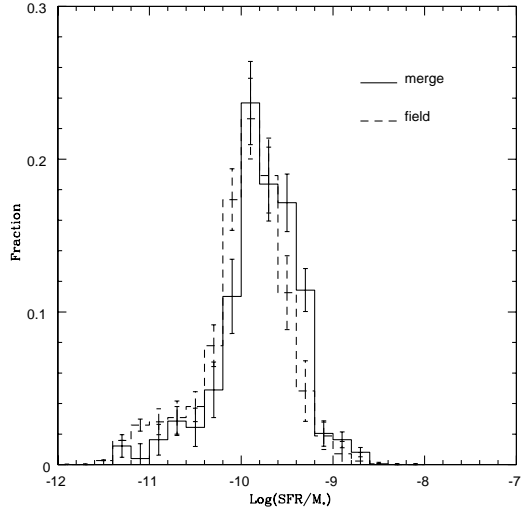}
  \end{center}
  \caption{The
distributions of spectroscopic indicators SFR/$M_{*}$ for galaxies
in the LGS (solid line) and its corresponding CGS (dashed line),
respectively.}\label{fig:fig4}
\end{figure}

\newpage

\section{Summary and Discussions}

In the present paper, three spectroscopic indicators: the 4000-\r{A}
break strength D$_{n}(4000)$, the Balmer absorption-line index
H$\delta_{A}$ and the specific star formation rate SFR/$M_{*}$, are
adopted to study the differences of star formation history between
merging galaxies and field galaxies. Our results show that, for
early-type merging galaxies, the interactions will not
\textbf{induce} significant enhancement of the star-forming
activities because of their stability and lack of cool gas. And for
late-type merging galaxies, they generally display more active star
forming activities than field galaxies on different timescales
within about 1Gyr. According to numerical simulations(\cite{tt72};
\cite{mh96}), the interaction between disk galaxies with comparable
masses can initiate a strong inflow of gas and trigger intense but
short-lived bursts of star formation. So the enhancement of
star-formation in our LGC could be attributed to gravitational
torques induced by galaxies interaction.

Since the strengthes of the 4000-\r{A} break and the H$\delta_{A}$
absorption line index can constrain the ages of the stellar
populations in galaxies and are able to distinguish recent star
formation histories dominated by bursts from those that are more
likely continuous(\cite{kh31}), the smaller D$_{n}(4000)$ and larger
H$\delta_{A}$ in our merging sample imply that the mean stellar age
of merging galaxies is younger than that of field galaxies and the
fraction of stellar masses formed during bursts over the past few
Gyr in merging galaxies is larger than field galaxies, which has
been further confirmed by the distributions of SFR/$M_{*}$ for
galaxies in the LGS and its corresponding CGS respectively.

Finally, it should be pointed out that
(about 65\% merging galaxies) in the LGC do not show the significant
enhancement of the star-formation. It could be due to the
complicated processes of star formation, especially for merging
galaxies, such as the internal structure of the merging galaxies,
merging orbital geometry and the rate of mass between merging pairs
(i.e. major or minor merging) etc. In order to determine accurately
how these factors work on the characters of merging galaxies, larger
samples are needed for further investigations.
\\

Funding for the creation and distribution of the SDSS Archive has
been provided by the Alfred P. Sloan Foundation, the Participating
Institutions, the National Aeronautics and Space Administration, the
National Science Foundation, the U.S. Department of Energy, the
Japanese Monbukagakusho, and the Max Planck Society. The SDSS Web
site is http://www.sdss.org/.

The SDSS is managed by the Astrophysical Research Consortium (ARC)
for the Participating Institutions. The Participating Institutions
are The University of Chicago, Fermilab, the Institute for Advanced
Study, the Japan Participation Group, The Johns Hopkins University,
the Korean Scientist Group, Los Alamos National Laboratory, the
Max-Planck-Institute for Astronomy (MPIA), the Max-Planck-Institute
for Astrophysics (MPA), New Mexico State University, University of
Pittsburgh, University of Portsmouth, Princeton University, the
United States Naval Observatory, and the University of Washington.

We thank Guinevere Kauffmann for helpful discussions and comments.
This work is partly supported by the NSFC Projects (No.
10503001,10403008,10333020 \& 10528307) and Science Foundation of
Shanghai No. 03XD14014 \& 05DZ22314.

\end{document}